\documentclass[fleqn,usenatbib]{mnras}
\usepackage{subfigure}
\usepackage{rotating}
\usepackage{newtxtext,newtxmath}
\usepackage[T1]{fontenc}
\usepackage{ae,aecompl} 
\usepackage{tikz}
\usepackage{amsmath}
\usepackage{nccmath}
\usepackage{multicol, blindtext}

\usepackage{graphicx}	
\usepackage{amsmath}	
\usepackage{float}		
\usepackage{placeins}
\usepackage{array}
\usepackage{tabularx}
\usepackage{gensymb}
\usepackage{soul}

\usepackage{newtxtext,newtxmath}

\usepackage[T1]{fontenc}

\DeclareRobustCommand{\VAN}[3]{#2}
\let\VANthebibliography\thebibliography
\def\thebibliography{\DeclareRobustCommand{\VAN}[3]{##3}\VANthebibliography}

\usepackage{graphicx}	
\usepackage{amsmath}	

\title[Magnetic fields in star forming environments]
{Magnetic fields in star forming environments: how does field strength affect gas on spiral arm and cloud scales?}

\author[N. P. Herrington et al.]{
Nicholas P. Herrington,$^{1}$\thanks{E-mail: nicholas.herrington.astro@gmail.com}
Clare L. Dobbs,$^{1}$
Thomas J. R. Bending$^{1}$
\\
$^{1}$School of Physics and Astronomy, University of Exeter, Stocker Road, Exeter, EX4 4QL, UK}

\date{Accepted XXX. Received YYY; in original form ZZZ}

\pubyear{2024}

\begin{document}
\label{firstpage}
\pagerange{\pageref{firstpage}--\pageref{lastpage}}
\maketitle

\begin{abstract}
We investigate star formation from subparsec to kpc scales with magnetohydrodynamic (MHD) models of a cloud structure and a section of galactic spiral arm. We aim to understand how magnetic fields affect star formation, cloud formation and how feedback couples with magnetic fields on scales of clouds and clumps.
We find that magnetic fields overall suppress star formation 
by $\sim$10\% with a weak field (5 $\mu$G), and $\sim50$\% with a stronger field (50 $\mu$G). 
Cluster masses are reduced by about 40\% with a strong field but show little change with a weak field.
We find that clouds tend to be aligned parallel to the field with a weak field, and become perpendicularly aligned with a stronger field,
whereas on clump scales the alignment is more random. 
The magnetic fields and densities of clouds and clumps in our models agree with the Zeeman measurements of the Crutcher relation $B-\rho$ in the weaker field models, whilst the strongest field models show a relation which is too flat compared to the observations.
In all our models, we find both subcritical and supercritical clouds and clumps are present.
We also find that if using a line of sight (1D) measure of the magnetic field to determine the critical parameter, the magnetic field, and thereby also criticality, can vary by a factor of 3-4 depending on whether the direction the field is measured along corresponds to the direction of the ordered component of the magnetic field. 
\end{abstract}




\section{Introduction}
Magnetic fields permeate the interstellar medium (ISM) making them important on all scales of the star formation process, from molecular clouds down to the protostellar environment. 
Observations indicate that the median magnetic field strength in the cold neutral medium (CNM) of the ISM is $\sim6~\mu$G \citep{crutcher1999,heiles2005,crutcher2010}, but can be as high as $20-30~\mu$G in M51 and $50-100~\mu$G in M87 (see \citealt{beck2015} and references therein). At these field strengths, magnetic energy densities are comparable  
to kinetic energy densities arising from turbulence, but smaller than thermal energies except for the hottest phases of the ISM \citep{crutcher1999,cox2005,beck2015}. Thus magnetic fields could be expected to be as significant as other processes in the ISM such as turbulence and feedback, and potentially provide support against gravitational collapse on core or cloud scales. 

The magnetic criticality of a cloud is a measure of whether a cloud will be supported against collapse by magnetic fields or not. The magnetic criticality can be computed by comparing a cloud's mass to its critical mass $M_{\scriptscriptstyle\textrm{crit}}$.  
The collapse of a cloud is expected to proceed if its mass to flux ratio exceeds a critical value $M/M_{\scriptscriptstyle\textrm{crit}} = \alpha_B \Phi_{\scriptscriptstyle B}/G^{1/2}$, where $\Phi_{\scriptscriptstyle B} = B\pi R^2$ \citep{mouschovias1976,nakano1978,krumholz2011} and $\alpha_B$ is $\sim$0.17 for a uniform spherical cloud. This can alternatively be expressed in terms of observable quantities, e.g. $M/M_{\scriptscriptstyle\textrm{crit}}\sim7.6\times10^{-21}N(H_{\scriptscriptstyle 2})/B_{\scriptscriptstyle\textrm{TOT}}$, where $N$ is the column density and $B_{\scriptscriptstyle\textrm{TOT}}$ is the magnetic field strength \citep{crutcher2004}. A structure for which $M/M_{\scriptscriptstyle\textrm{crit}}>1$ is referred to as supercritical. For mass to flux ratios below the critical value (i.e. $M/M_{\scriptscriptstyle\textrm{crit}}<1$) structures are subcritical, i.e. magnetically supported against collapse. We know that clouds are evolving structures and so their criticality may not stay consistent throughout their lifetimes. 
This has lead to two different pictures of magnetic fields and their role in the star formation process, referred to as strong field and weak field models \citep{shu1987,crutcher2012}. In strong field scenario clouds are born subcritical, collapsing only once they become supercritical through accretion of mass increasing their $M_{\scriptscriptstyle\textrm{crit}}$ \citep{vazquez-semadeni2011,kortgen2018}
and or through ambipolar diffusion of the constituent gas and ions (e.g., \citealt{parker1967,mouschovias1979,shu1983}).
Weak field models (e.g., \citealt{padoan1999,maclow2004}) predict that clouds are controlled by turbulent flows and either dissolve back into the interstellar medium (ISM) or collapse if self gravitating upon formation.  
\citet{vazquez-semadeni2011} looked at the role of turbulence and magnetic fields by evaluating criticality and star formation properties of clouds structures formed in converging flows. They find that the criticality of clouds is very sensitive to location within the cloud, suggesting regions of a cloud can be supercritical whilst globally magnetically supported. This was also found in work from \citet{hu2023}, as well as concluding that most clouds are supercritical and subcritical clouds are identified as such due to observational biases.

Observations highlight the dependency of magnetic field strength with density across scales from clouds to cores (see reviews by \citealt{crutcher2012} and \citealt{pattle2023}). Line of sight Zeeman measurements suggest that molecular clouds are mostly supercritical following a $B \propto n^{\kappa}$ relation where $\kappa \approx 2/3$ \citep{crutcher2010}. Theoretically, there are a number of ways $B$ can be expected to depend on $n$ \citep{crutcher2012}. 
The most extreme cases are when gas is compressed along ($\kappa=0$ as the field does not change) or perpendicular ($\kappa=1$) to the field. For the weak field case, there is no preferred direction and $\kappa\approx 2/3$ \citep{mestel1996}. When the field is strong and ambipolar diffusion is acting to enable compression, $\kappa \leq 0.5$ \citep{mouschovias1981,mestel1984,mouschovias1999,auddy2022}.
Observations also indicate a transition from a flat relation at low densities, to a power law relation at high densities, the transition appearing to occur at densities of $n\sim100-1000$ cm$^{-3}$.
The transition is usually assumed to coincide with a change from structures being subcritical to supercritical \citep{crutcher2012}. 
Models of a turbulent magnetised ISM by \citet{auddy2022} including ambipolar diffusion also find the transition for the $B - \rho$ occurs at a density $\rho_{\scriptscriptstyle T}$ related to the initial Alfvén Mach number, $\mathcal{M}_{\scriptscriptstyle\textrm{A0}}$. 

Magnetic fields are likely to effect the structure of the interstellar medium on different scales, and ultimately the star formation rate. Magnetic fields provide an additional pressure, on large scales inhibiting the formation of dense structures such as clouds and interarm spurs \citep{dobbsp2008}. \citet{wibking2023} find the addition of magnetic fields leads to a factor of 1.5-2 decrease in the star formation rates on a galaxy scale. Other studies which follow the growth of galactic magnetic fields identify a reduction in the star formation rate once the magnetic field amplification saturates in the cold phase (e.g. \citealt{wanga2009,pakmor2013,hennebelle2014,vanloo2015}). On molecular cloud scales, simulations similarly find reduced star formation rates (e.g. \citealt{priceb2009,Padoan2011,peters2011,federrath2012,federrath2015,kortgen2015,vanloo2015}). 
At the level of cores, the addition of magnetic fields is found to reduce fragmentation, leading to the formation of fewer brown dwarfs and producing an Initial Mass Function (IMF) in better agreement with observations  
\citep{priceb2008}.
An example of an exception to this trend is a colliding flow study from \citet{zamora-aviles2018}. They find magnetic fields aligned to inflows result in the formation of less turbulent cloud structures which more readily collapse and form stars compared to purely hydrodynamic case. 

The direction of the magnetic field in the ISM is a further observational measure of the influence of the magnetic field on gas. Filamentary structures within the diffuse ISM appear to be aligned either parallel (at lower column densities) or perpendicular (at higher column densities) to their local magnetic field \citep{mccluregriffiths2006,planckcolab2016,solera2017,alina2019,lee2021}. Simulations often find the transition between parallel and perpendicular alignments occurs with increasing magnetic field strength and/or increasing density (e.g. \citealt{soler2013,klessen2017,xu2019,seifried2020,dobbsw2021,barretomota2021,mazzei2023}). \citet{girichidis2021} additionally identifies this transition and finds that it coincides with a change in the angle between the flow and the magnetic field (i.e. parallel or perpendicular).
The transition from parallel to perpendicular orientations has been related to shocks from colliding streams of gas \citep{inoue2009,inoue2016,dobbsw2021}, strongly converging flows in turbulent clouds \citep{solerh2017}, and the transition from magnetically supported to gravitationally dominated filaments \citep{solerh2017,girichidis2021}.

Most studies to date focus on either molecular cloud or whole galaxy scales. In this study we investigate the evolution of a kpc scale star forming region, consisting of a section of galactic spiral arm extracted from a pre-evolved galaxy model. We additionally investigate a 100 pc scale cloud structure, similarly extracted from a pre-evolved galaxy model. These models include magnetohydrodynamics (MHD) and stellar feedback. We aim to understand the magnetic criticality of cloud and clump structures, the orientation of filaments with respect to elongation to their magnetic fields and how magnetic fields interact with stellar feedback. This paper is structured in the following way: in Section \ref{sec:method} we describe our simulation setup, in Section \ref{sec:results} we show the evolution of our Arm and Cloud scale models, the magnetic properties of cloud and clump structures, the magnetic field density relation, the interaction between feedback and magnetic fields and lastly the properties of cluster in our pc scale models. In Section \ref{sec:disc} we discuss methods to calculate the critical mass to flux ratio, a discussion of magnetic fields from the context of our models. A summary of our main results is given in Section \ref{sec:conclusions}.

\section{Methodology}
\label{sec:method}
Our simulations are modelled using the magnetohydrodynamics code sphNG \citep{benz1990,benzetal1990,batebonprice1995,pricemon2007}, featuring ISM chemistry for heating and cooling \citep{glover2007}, magnetic fields \citep{price2012} and non ideal MHD \citep{wurster2014,wurster2016}), sink particles \citep{bateb1995} and self gravity. Shocks are captured via artificial viscosity \citep{Morris1997}, where the viscosity parameter can vary between 0.1 and 1. The default smooth particle magnetohydrodynamics parameters for artificial resistivity are used, i.e. $\alpha_{\scriptscriptstyle B} = 1$. The code is parallelised with both OpenMP and message passing interface (MPI).
\subsection{Magnetohydrodynamics}
The simulations in this paper all assume ideal MHD. To keep the magnetic divergence close to zero, we use a
hyperbolic divergence cleaning method 
\citep{Dedner2002}. This method couples the magnetic field to a scalar field, through which the divergence error can be subtracted by propagating $\nabla \cdot \textbf{B}$ as a damped wave.  We use fast MHD waves with wave speed $c_{\scriptscriptstyle h}$ for the divergence cleaning, which is sufficient to enforce $h\nabla\cdot\textbf{B}/|\textbf{B}|<0.01$. The cleaning is controlled by an over-cleaning parameter which we set as $\sigma=1$, the default value.
The magnetic field evolution follows the quantity $\textbf{B}/\rho$, with the magnetic tension instability handled by the \citet{borve2001} method, see \citet{price2012} for further details.

\begin{table}
\caption{List of models, their initial magnetic field strengths, simulation lifetime, progenitor galaxy model and mass resolution. G1 and G2 refer to the galaxy models from \citet{DobbsPringSpiral2013} and \citet{pettitt2015} respectively.}
\resizebox{\columnwidth}{!}{\begin{tabular}{| c | c | c | c | c |}
\hline
 Model & Initial magnetic & Simulation  & Galaxy & Mass resolution\\
 Name & field strength & time (Myrs) & progenitor & (M$_{\odot}$)\\
 & ($\mu G$)&&(G1/G2)&\\[1ex]
 \hline
 Arm1 & 1 & 5.6& &\\ \cline{1-3}
 Arm5 & 5 &6.8&G1 &3.67\\ \cline{1-3}
 Arm10 & 10 &5.5 & &\\ \cline{1-3}
 Arm0 & -& 5.8& &\\ \cline{1-5}
 Cloud5 & 5 & 6.3 & &\\ \cline{1-3}
 Cloud50 & 50 & 7.1 &G2 &0.02\\ \cline{1-3}
 Cloud0 & - & 6.0 & &\\ 
 \hline   
\end{tabular}}
\label{tab:models}
\end{table}
\begin{figure*}
    \centering
    \includegraphics[width=170mm]{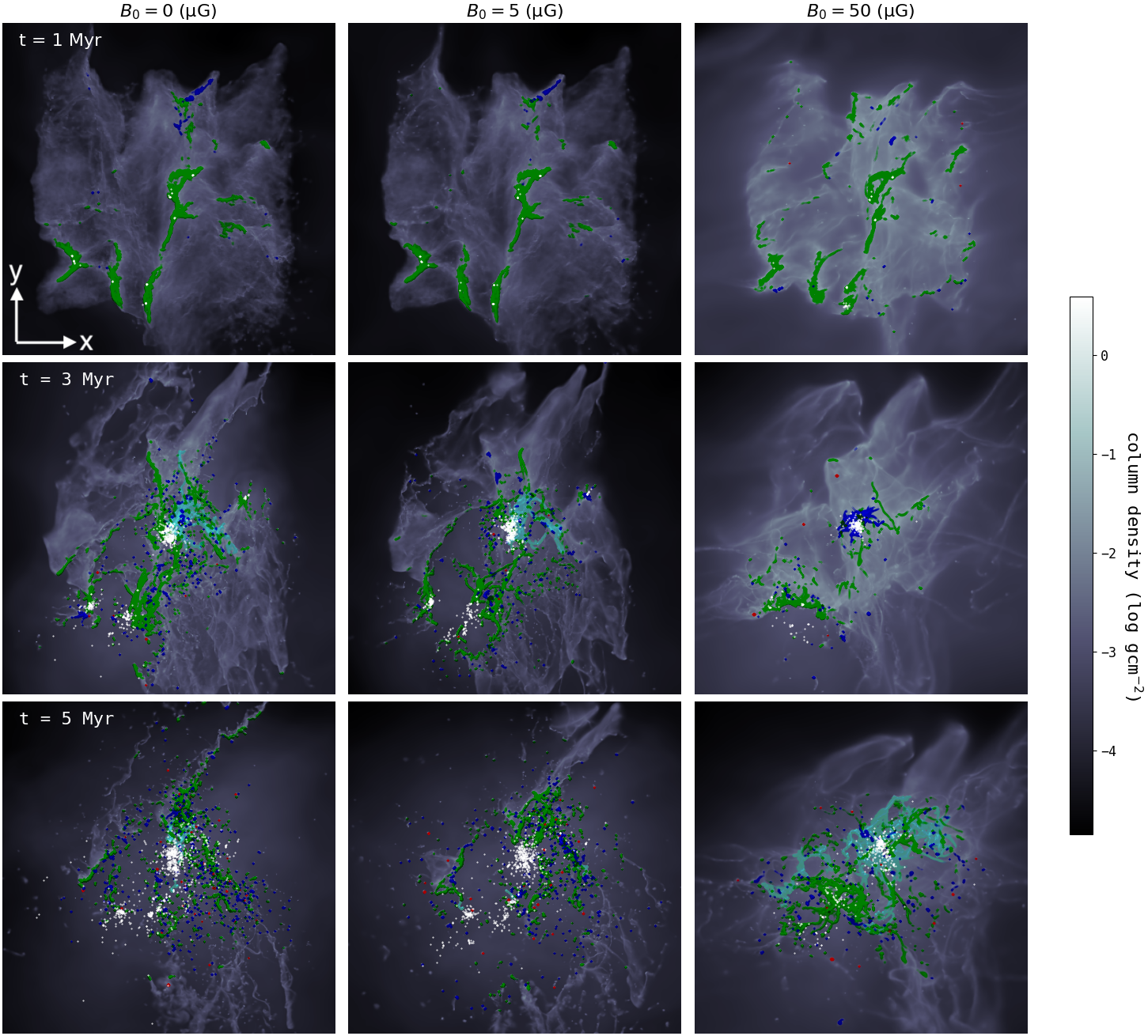}
    \caption{Column density snapshots of Cloud0, Cloud5, Cloud50 at times of 1, 3 and 5 Myrs. Clumps are overlayed in colour based on their morphology. Green clumps are filamentary, blue clumps are sheet like, turquoise are curved sheets and red clumps are spheroidal. Sink particles are plotted in white.}
    \label{fig:cd_all}
\end{figure*}
\begin{figure*}
    \centering
\includegraphics[width=170mm]{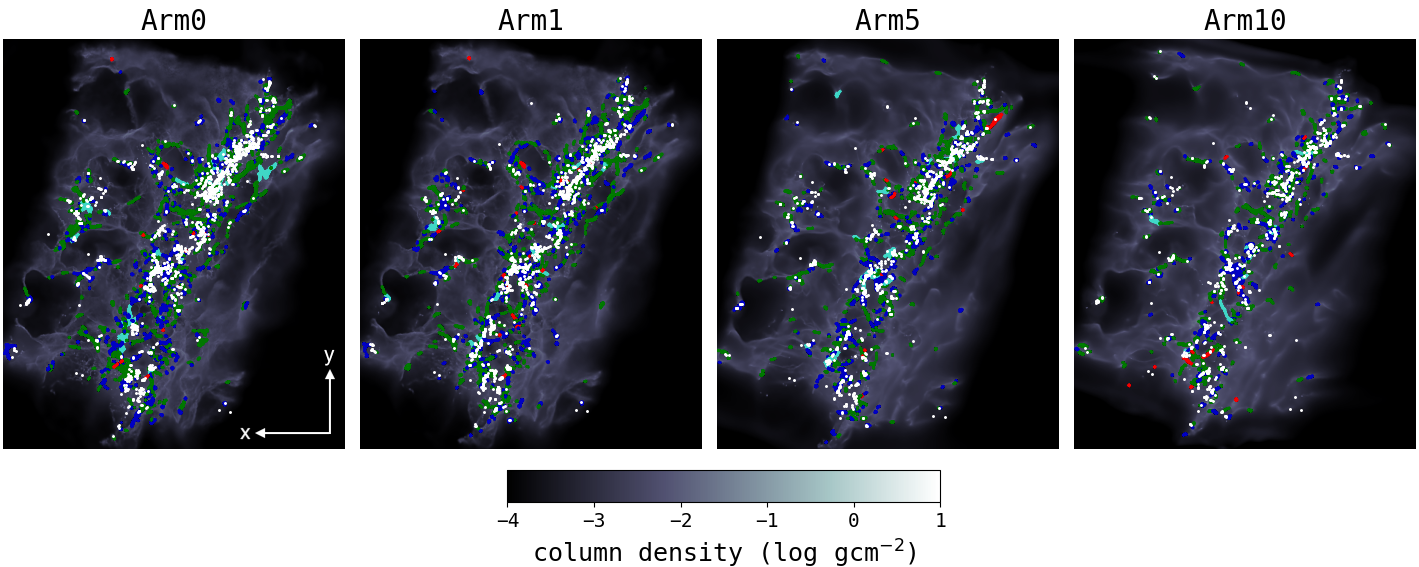}
    \caption{Column density snapshots of models Arm0, Arm1, Arm5 and Arm10 at 6 Myrs. White points are cluster sink particles, green are filaments, blue are sheets, red spheroidal and turquoise are curved sheets.}
    \label{fig:cd_all2}
\end{figure*}
\subsection{Simulation details}
The simulations performed here largely use the same setup as those presented in \citet{Herrington2023} and \citet{dobbs2022}, but now focus on magnetic fields. Our simulations based on \citet{Herrington2023} follow a kpc size region and contain an initial population of stars. 
The initial conditions are taken from a larger scale galaxy simulation (see Section 2.3), including sink particles which contain a population of stars. Stars are assigned to these sinks, using a method described later in this section. We refer to these as `Arm' models. Simulations based on \citet{dobbs2022} follow a smaller, 100 pc region without an initial population and are referred to as `Cloud' models.
In all the models, we also insert sink particles 
when gas exceeds given densities, following \citet{Bate1995sinks}. 
The sinks represent small clusters of stars and so can also be thought of as cluster sinks.
Cluster sink particles are inserted when gas exceeds densities of 3000 cm$
^{-3}$ and 1000 cm$^{-3}$ in the Arm and Cloud models respectively, and can accrete gas with accretion radii of 0.6 and 0.1 pc. 

Stellar feedback from massive stars is modelled by a prescription for photoionising and supernovae feedback. For photoionisation a line of sight (LOS) approach is used as described in \citet{Bending2020}, in which LOS are drawn for each gas particle around an ionising source. Along each LOS the column density contribution of intersecting particles is added to the total column density. The on-the-spot approximation is used for emission of 13.6 eV photons through recombination, with a recombination rate of $2.7 \times 10^{-13}$~cm$^3$s$^{-1}$. Photoionised gas is heated to $10^{4}$~K and can begin cooling once its ionised fraction drops below 1\%.
Supernovae (SNe) are inserted using the same method as \citet{dobbs2011_sn}. The ambient gas density is calculated from the neighbouring gas particles, which is then used to calculate the velocity and temperature to be assigned to the gas, assuming one supernovae deposits $10^{51}$ ergs of energy.
The injected SNe represent 
the pressure-driven snowplough phase, since the prior free expansion and adiabatic phases are not necessarily well resolved.

As in previous work, we use a star formation efficiency (SFE) parameter referring to the fraction of a cluster sink's mass that we assume to be star mass. This parameter is 0.25 in the Arm models, and 0.5 in the Cloud models. We allocate stars to the sink particles using a pre-computed sample of stars \citep{Bending2020}. They are inserted into sinks throughout the evolution of both models as well as into the initial population of sinks in the Arm models. The pre-computed sample has an associated massive star injection interval, $\Delta M_i$, of 305~M$_{\odot}$ for this work. This value is calculated by dividing the total stellar mass of the sample by the 
number of stars above 8~M$_\odot$. 
Massive star insertion occurs each time the total sink mass multiplied by the SFE parameter increases by $\Delta M_i$. For each cluster sink particle we find the target star mass, defined as the particle mass multiplied by the SFE parameter. We also find the current star mass, defined as the number of massive stars already added to the sink multiplied by $\Delta M_i$. We then subtract the current mass from the target mass for each cluster sink particle. If the sum of these differences is greater than $\Delta M_i$ we add another massive star to the cluster sink particle with the maximum difference.
The lifetimes and stellar fluxes of the massive stars are calculated using the stellar evolution program SEBA \citep{Portegies2012}. Only stars with masses > 18~M$_{\odot}$ are considered to be ionising, 
as the ionising output from the lower mass end of the sample is negligible. Supernovae occur for all stars > 8 M$_{\odot}$ once they have reached their lifetime within the simulation. 

\subsection{Initial conditions}
\label{section:ICs}
We present models on two differing scales, which as discussed in the previous section we split into `Arm' and `Cloud' models. Arm models are split into four simulations, three MHD and one HD model. Cloud models are split into three, with two MHD and one HD model. The initial magnetic field strengths of each MHD model are shown in Table \ref{tab:models}. We label our models based on their initial magnetic field strength, i.e. $B_0 = 0, 1, 5$ and $10~\mu$G corresponds to Arm0, Arm1, Arm5 and Arm10. Our Cloud models follow the same naming convention. The initial conditions for both the Arm and Cloud models are extracted from pre-evolved galaxy models.

The initial conditions of our Arm models are setup as a combination of IOSFB and MHD5 from \citet{Herrington2023}, in which we extracted a 1.2 kpc$^3$ box from a galaxy evolution model from \citet{Dobbs2017} and apply a clockwise toroidal magnetic field across the region. The region is in the upper left quadrant of the galaxy, the initial magnetic field geometry approximately follows the spiral arm and so is pointing predominantly along the y-axis. The galaxy model included ISM heating and cooling, self gravity and a prescription for supernovae feedback. Once we extract the gas from the galaxy we perform a resolution increase of the same method as \citet{Bending2020}, to improve our mass per particle resolution from 312.5~M$_{\odot}$ to 3.68~M$_{\odot}$. 
We additionally ran a couple of lower resolution simulations of the Arm5 model, to test the effect of resolution, and the modelling of turbulence on the smallest scales in our simulations. We discuss these simulations in Section \ref{sec:magcrit}, Section \ref{sec:conclusions} and Appendix \ref{append:sim_convergence}. As well as this, we discuss the impact of our initially uniform magnetic field distribution in Section \ref{sec:conclusions} and 
Appendix \ref{append:initial_mtof}.

The second set of initial conditions for our Cloud models are based on the M1R1FB simulation from \citet{dobbsb2022}, which models a region of around 100~pc size scale. The initial conditions were similarly extracted from a galaxy scale simulation from \citet{pettitt2015} using the same resolution increase scheme. Again a toriodal magnetic field is applied, with initial field strengths of 5 and 50~$\mu$G, 
as the region is in the upper right quadrant of the galaxy
the initial magnetic field geometry points in a south-eastern direction.

Similarly to \citet{wurster2021} we include a surrounding isothermal boundary which represents a warm ionised medium. Particles are placed within a volume around the region such that they are in pressure equilibrium with the extracted region. These boundary particles are fixed to a temperature of $2\times10^4$~K. We set the velocity of a boundary particle to match that of the closest particle from the extracted region. This boundary medium is included in both the Arm and Cloud scale models.

For both the Arm and Cloud models, the simulations differ from previous simulations due both to the addition of the magnetic field, and the boundary conditions. Because the addition of the boundary conditions mean the initial conditions are not the same even without magnetic fields, non-MHD calculations were also carried out for comparison. 

\begin{figure}
    \centering
    \includegraphics[width=68mm]{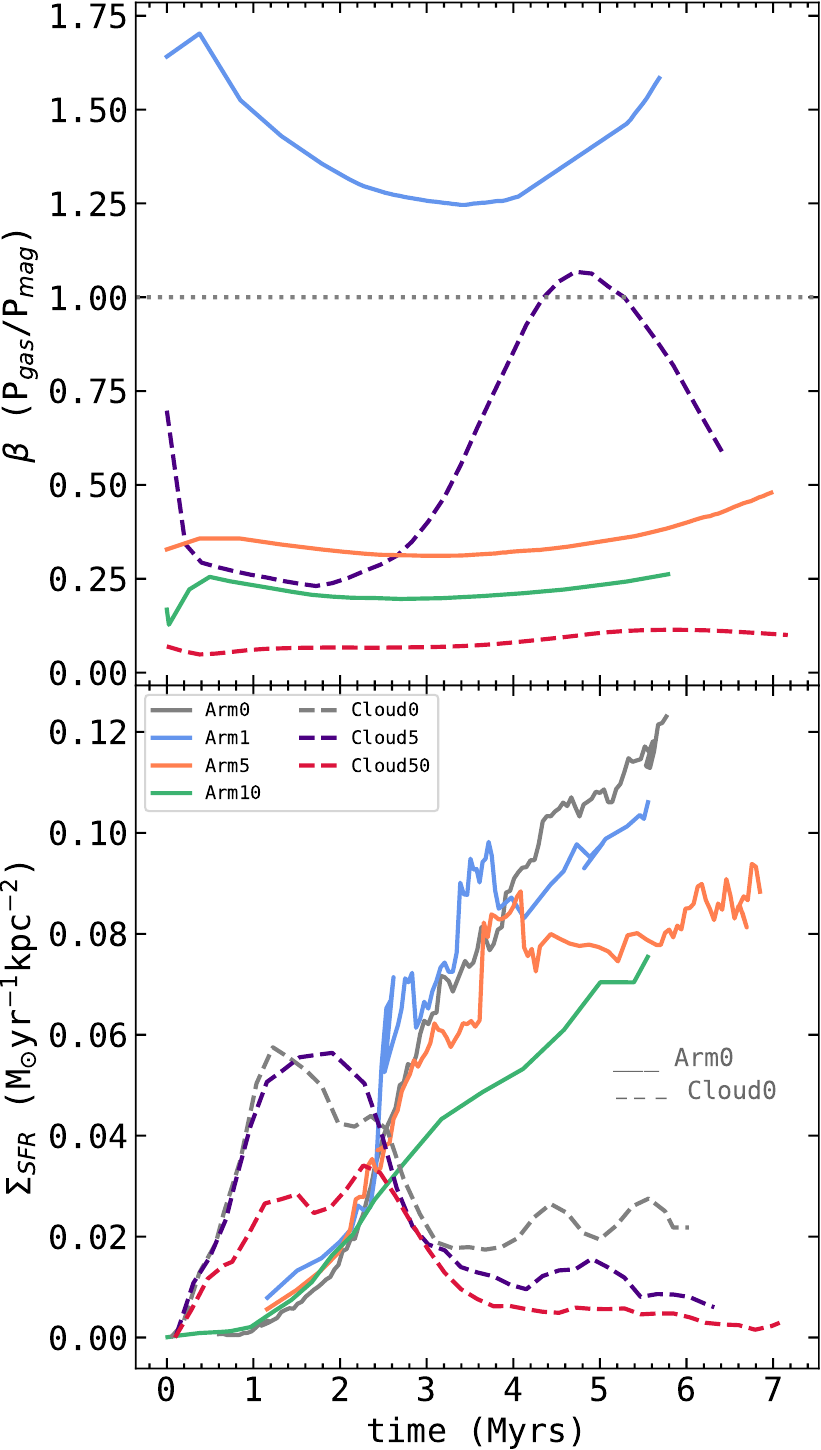}
    \caption{
    The top panel shows the ratio of gas pressure to magnetic pressure, plasma $\beta$, in which the dashed line indicates $P_{\scriptscriptstyle\textrm{gas}} = P_{\scriptscriptstyle\textrm{mag}}$. The lower panel shows the star formation rate surface density for hydrodynamical and MHD models. Hydrodynamical models are shown as grey lines.}
    \label{fig:props_gt}
\end{figure}

\section{Results}
\label{sec:results}
\subsection{Evolution of Arm and Cloud models}
\begin{figure*}
    \centering
    \includegraphics[width=165mm]{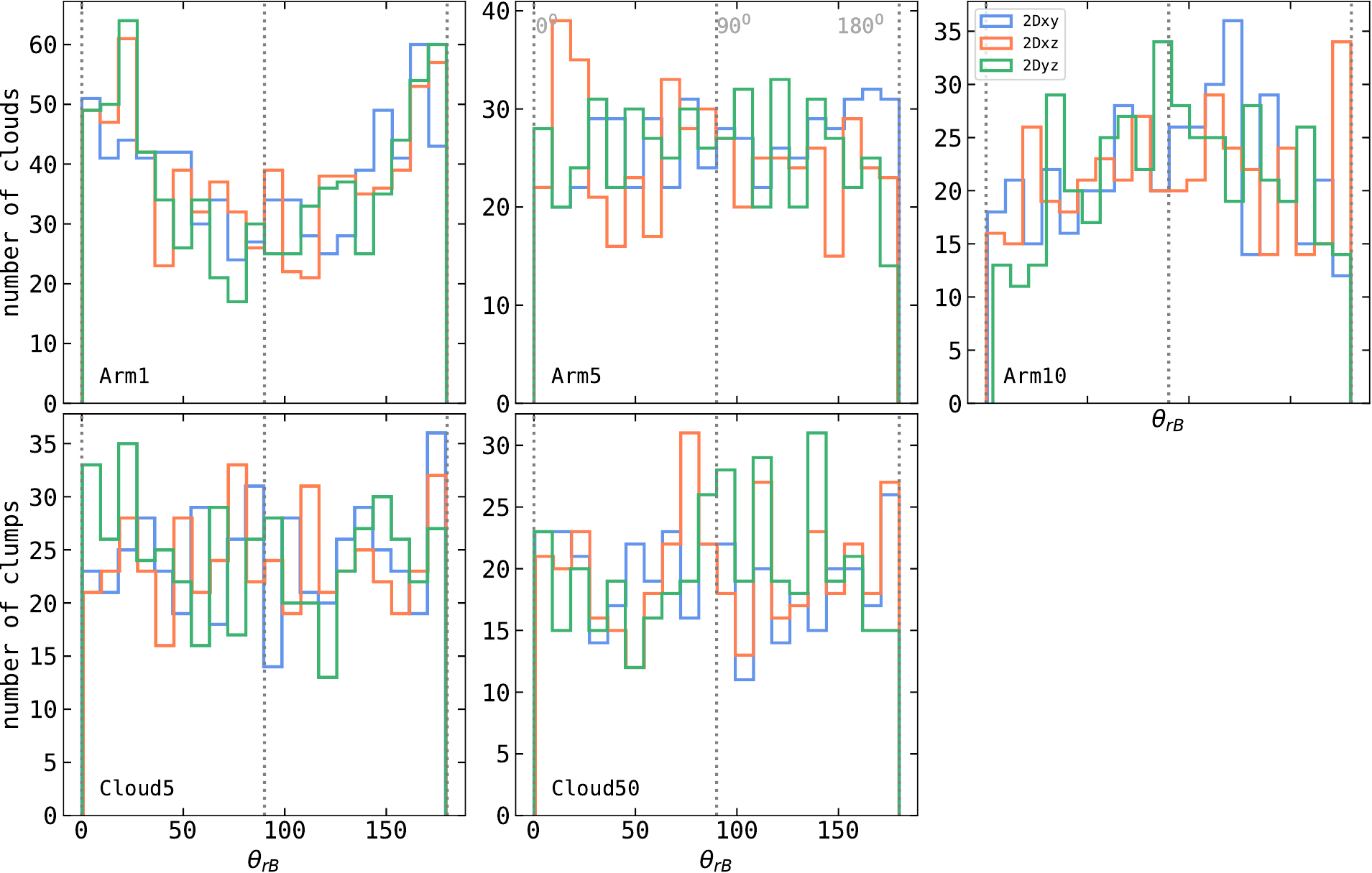}
    \caption{Histograms of the angle between a filament's long axis and its internal magnetic field, $\theta_{\scriptscriptstyle \textrm{rB}}$, for all Arm and Cloud MHD models. These angles are calculated in 2D, including three slices: 2D$_{\scriptscriptstyle\textrm{xy}}$, 2D$_{\scriptscriptstyle\textrm{xz}}$ and 2D$_{\scriptscriptstyle\textrm{yz}}$, computed for each model. $\theta_{\scriptscriptstyle rB}$ can range between 0$^{\scriptscriptstyle\circ}$ and 180$^{\scriptscriptstyle\circ}$, with angles close to 0$^{\scriptscriptstyle\circ}$ indicating parallel and close to 180$^{\scriptscriptstyle\circ}$ being anti-parallel.}
    \label{fig:angles_clumps}
\end{figure*}
We show the evolution of the cloud scale models Cloud0, Cloud5 and Cloud50, at times of 1, 3 and 5~Myr, in Figure \ref{fig:cd_all}.
Comparing the control model (Cloud0) to the MHD runs Cloud5 and Cloud50 we see that increasing the field strength leads to much smoother structures. At 3~Myr, Cloud0 and Cloud5 are similar in structure, displaying  feedback bubbles produced by ionising massive stars. In Cloud50 however these same feedback bubbles are not apparent, and the gas remains relatively smooth and undisturbed. By 5~Myr both Cloud0 and Cloud5 show significant gas dispersal from the 
ionising massive stars 
in the clouds. Within Cloud50 we start to see bubbles swept out by feedback, but there is comparatively less gas dispersal. In all models we observe 
a large central cluster formed from the collapse of the main filamentary structures threading the middle of the clouds.
The population of stars within Cloud0 and Cloud5 form around 3 further small but distinct clusters after 5~Myrs, but we only clearly see the one central large cluster in Cloud50.  
\begin{figure*}
    \centering
    \includegraphics[width=130mm]{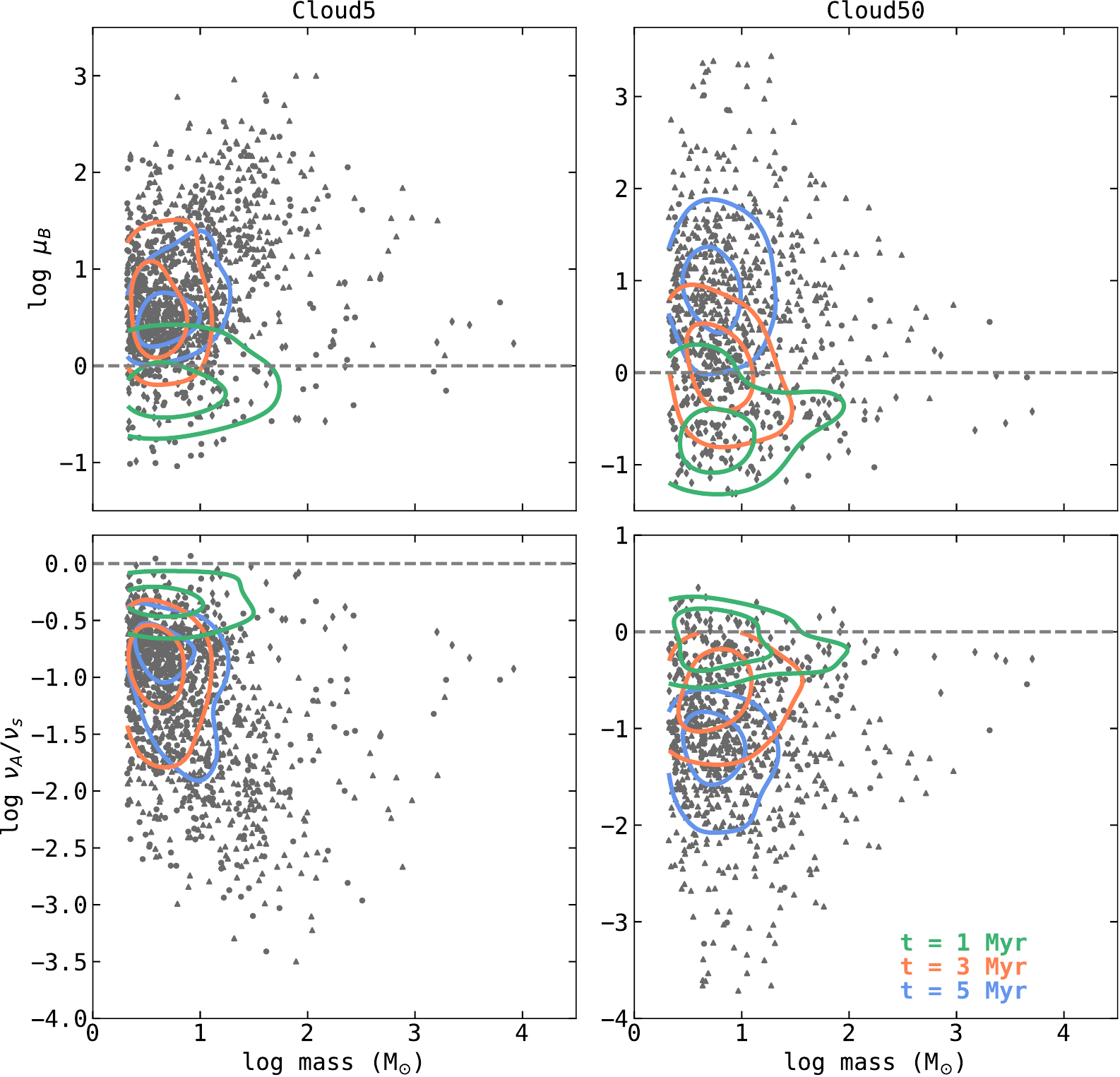}
    \caption{The mass to flux ratio, $\mu_{\scriptscriptstyle B}$, is shown in the top row and the ratio of the Alfvén speed to the sound speed, $\nu_{\scriptscriptstyle A}/\nu_{\scriptscriptstyle s}$, is shown in the bottom row. These are calculated for clumps in Cloud5 and Cloud50 models at times 1, 3 and 5~Myrs. The initial mass to flux ratios of Cloud5 and Cloud50 are also plotted, marked as a cross. The green contours correspond to the diamond points at t = 1~Myrs. Circle points are associated with the orange contours at t = 3~Myrs. Triangle points represented by the blue contours refer to times t = 5~Myrs. The contours are created using a Gaussian Kernel Density Estimate (KDE) function where the bandwith is calculated using Scott's rule.}
    \label{fig:props_clumps}
\end{figure*}

We show column density snapshots of our arm regions Arm0, Arm1, Arm5 and Arm10 at 6~Myr in Figure \ref{fig:cd_all2}. Similarly to the cloud models, we see the gas structure are increasingly smoothed with increasing initial magnetic field strength in the arm models.
Feedback bubbles are smoother and less structured in the higher field strength models. Filamentary and sheet like clouds that form around feedback bubbles are compressed and fragmented in the magnetic field models, with increasing fragmentation 
occurring with increasing field strength. In our non magnetic model Arm0 we see the largest number of 
sink particles and cloud structures across the spiral and inter arm regions. With increasing field strength, we see fewer sinks and cloud structures, particularly in the inter-arm regions.

In Figure \ref{fig:props_gt}, we show the evolution of the ratio of the gas pressure to magnetic pressure (plasma $\beta$) and the star formation rate surface density ($\Sigma_{\scriptscriptstyle \textrm{SFR}}$) for all models. 
Initially, the plasma $\beta$ parameter (second panel) for  Cloud5 decreases, indicating that Cloud5 becomes more magnetically dominant, but this is then reversed 
and Cloud5 becomes more gas pressure dominant after a few million years. This is because the density increases as collapse occurs within the cloud, and also as feedback occurs the temperature increases. Similar, fluctuating behaviour is seen for Arm1, but for the higher strength models, where the magnetic pressure is higher, there is less change in the densities compared to the initial conditions.
Cloud50 remains dominated by magnetic pressure ($\beta \lesssim 0.1$) during its evolution,
and likewise magnetic pressure dominates for the Arm5 and Arm10 models. In all these magnetically dominated models there is little variation in $\beta$. 

The amount of star formation is larger in Cloud5 than Cloud50, with the star formation rate roughly double 
at all times. We expect the reduced star formation in Cloud50 results in less feedback and the reduced level of gas dispersal that we see in Figure \ref{fig:cd_all}. The Arm models follow a similar trend to the Cloud models, i.e. a stronger initial magnetic field strength results in 
lower star formation rates. Similarly to models in \citet{Herrington2023} the initial star formation rates in our new arm models are roughly equal during the first 2 million years. After this the magnetic models Arm5 and Arm10 begin to plateau at lower star formation rates compared to Arm0. 
Arm0 and Arm1 exhibit similar star formation rates
up to 4~Myr before 
the star formation rate is suppressed
in Arm1.

\subsection{Clump and Cloud properties}
\label{sec:cl_props}
\subsubsection{Morphologies}
\label{sec:morphs}
\begin{figure*}
    \centering
    \includegraphics[width=165mm]{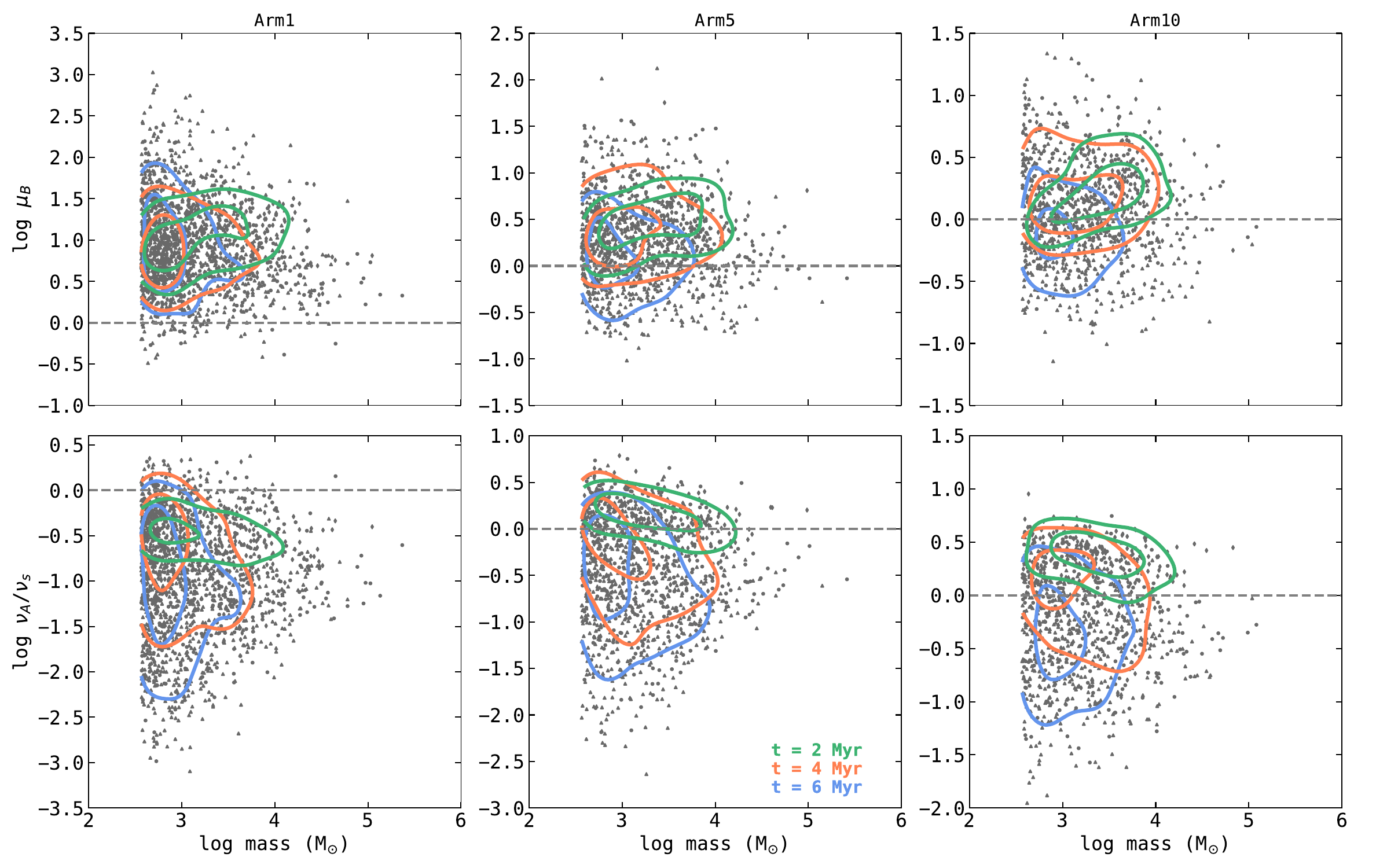}
    \caption{The mass to flux ratio, $\mu_{\scriptscriptstyle B}$, is plotted in the top row and the ratio of the Alfvén speed to the sound speed, $\nu_{\scriptscriptstyle A}/\nu_{\scriptscriptstyle s}$, in the bottom row. These are calculated for clouds in Arm1, Arm5 and Arm10 models at times 2, 4 and 6~Myrs. The green contours correspond to the diamond points at t = 2~Myrs. Circle points are associated with the orange contours at t = 4~Myrs. Triangle points represented by the blue contours refer to times t = 6~Myrs. The contours are created using Gaussian KDE function where the bandwith is calculated using Scott's rule.}
    \label{fig:props_clouds}
\end{figure*}
\begin{table*}
\caption{Percentage of clouds that are classified as sheet, curved sheet, filament and spheroidal for Arm and Cloud models. These are calculated at times of 6~Myrs and 5~Myrs for the Arm and Cloud models respectively.}
\begin{center}
\begin{tabular}{| c | c | c | c | c |}
 \hline
 Model & Sheet (\%) & Curved sheet (\%) & Filament (\%) & Spheroidal (\%) \\ 
 \hline
 Arm0  & 44.13 & 1.635 & 52.79 & 1.442 \\
 \hline
 Arm1 & 42.32 & 1.289 & 53.71 & 2.685 \\
 \hline
 Arm5& 40.29 & 2.408 & 55.38 & 1.926 \\
 \hline
 Arm10& 41.20 & 1.547 & 54.35 & 2.901 \\
 \hline
 Cloud0& 43.12 & 0.2200 & 52.48 & 4.180 \\
 \hline
 Cloud5 & 40.12 & 0.4559 & 55.17 & 4.255 \\
 \hline
 Cloud50& 33.40 & 0.2024 & 64.17 & 2.227 \\
 \hline
\end{tabular}  
\label{tab:morphs}
\end{center}
\end{table*}
In order to identify clumps or clouds within the Cloud and Arm models, we use a friends of friends algorithm (see \citealt{Bending2020,Ali2022,Herrington2023}). The algorithm uses a particle separation to compare the distance between a chosen particle and a potential group member. If this distance is within the defined particle separation the particle is added to the group. 
This process is repeated for ungrouped particles until all particles in the model are 
checked. The particle separation is 0.8 pc for the Arm models and 0.02 pc for the Cloud models. These length scales impose a minimum density for identified clouds and clumps, which is equal to $n_{\scriptscriptstyle\textrm{H}}\sim200$~cm${^{-3}}$ and $n_{\scriptscriptstyle\textrm{H}}\sim2200$~cm${^{-3}}$ for the clouds (Arm models) and clumps (Cloud models) respectively. With clumps and clouds identified, we classify 
their shapes using the principle axes theorem. We solve the eigenvalue problem for the moment of inertia equation ($\mathbf{L}=\mathbf{I \omega}$) to obtain the principle axes of inertia for a given clump or cloud. This provides us with eigenvalues and vectors for the corresponding axes which we use to classify 
the cloud's / clump's morphology. 

In the same way as \cite{ganguly2023} we can sort the eigenvalues for the principle axes into $a, b$ and $c$, where $a > b > c$, and define four morphologies as follows 
\begin{equation*}
    \textrm{\textbf{Filament:}} \ \frac{a}{b}>f_{\textrm{ar}};
\end{equation*}
\begin{equation*}
    \textrm{\textbf{Sheet:}} \ \frac{a}{b} \leq f_{\textrm{ar}} \ \& \ \frac{a}{c}>f_{\textrm{ar}};
\end{equation*}
\begin{equation*}
    \textrm{\textbf{Curved sheet:} \ \textrm{sheet but the centre of mass is outside the object}};
\end{equation*}
\begin{equation*}
    \textrm{\textbf{Spheroidal:}} \ \frac{a}{c} \leq f_{\textrm{ar}}.
\end{equation*}
$f_{\textrm{ar}}$ is the aspect ratio factor which is set to 3. An example of each classification is shown in Figure~\ref{fig:appendix_classify}.
The clumps and clouds are also shown in Figure \ref{fig:cd_all}, coloured according to morphology. In Table \ref{tab:morphs}, we break down the cloud or clump classifications as a percentage of the total number of structures in each model. The most frequent morphology is filamentary, followed by sheet-like, followed by spheroidal.
Over time the percentage of sheets structures increases in all Cloud models. Cloud0 shows a roughly equal split between filaments and sheets, Cloud5 is more weighted towards filaments and Cloud50 shows significantly more filaments then sheets, thus a stronger magnetic field results in a higher fraction of filaments. This is 
similar to results from \citet{ganguly2023}, as they also report a split in sheet like and filamentary structures along with a very small percentage of spheroidal structures. The Arm models similarly show clouds evenly distributed between sheets and filaments, and again some tendency for filaments to become slightly more common with increasing field strength.
We also see that the number of curved sheets and spheroidals encompasses only around $\sim3\%$ of structures in both the Arm and Cloud models. 

\subsubsection{Orientation of magnetic field and filaments}
\label{sec:orientations}
Figure \ref{fig:angles_clumps} shows the most common angle, $\theta_{\scriptscriptstyle \textrm{rB}}$, between the long axis of a filament $r$ and its magnetic field $B$, for all Cloud and Arm models at times 5 and 6 Myrs respectively. We have taken three different 2D slices, 2D$_{\scriptscriptstyle\textrm{xy}}$, 2D$_{\scriptscriptstyle\textrm{xz}}$ and 2D$_{\scriptscriptstyle\textrm{yz}}$, which are analogues to a bird's eye
view of the mid-plane (xy) and looking through the mid-plane (xz and yz) 
\footnote{The distribution of angles between two random vectors in 3D is biased towards perpendicular, so even if the magnetic field has no relation to the cloud axis it will appear preferentially perpendicular \citep{solerh2017}}. In the lowest magnetic field model, Arm1, we observe that clouds are generally orientated parallel to the magnetic field, with the majority of angles $\theta_{\scriptscriptstyle\textrm{rB}}$ between 0$^{\circ}$ to 30$^{\circ}$ (parallel) and 160$^{\circ}$ to 180$^{\circ}$ (anti-parallel). 
This is true in all three 2D slices, indicating the same trend regardless of viewing position. 
With increasing field strength, we see relatively fewer parallel orientations. Arm5 shows no preferred alignment of the clouds and their magnetic fields across all three 2D slices. Arm10 however does contain more perpendicularly aligned clouds (in all three 2D slices) suggesting that perpendicular alignments are favoured in higher magnetic field strength clouds. There is less of a trend for the Cloud models, although there are still relatively more parallel aligned clouds in Cloud5, and more clouds which are perpendicularly or closer to perpendicularly aligned in Cloud50.

Other numerical studies find a transition from filaments tending to be aligned parallel to the magnetic field, to being aligned perpendicular to the magnetic field, at high densities and with sufficiently high magnetic field strengths \citep{solerh2017,seifried2020,dobbsw2021,mazzei2023}. There is also observational evidence of a transition with density \citep{planckcolab2016,solera2017,alina2019,lee2021}, though the distribution can be more random for high contrast filaments \citep{alina2019} or on small, core scales (\citealt{chene2020,sharma2022}). We see the Arm models agree with a dependence on the magnetic field. A further test, Arm50 (with $B=50~\mu$G) showed filaments preferentially perpendicular to both the field and the spiral arm compared to the other models. Both the Arm and Cloud models are quite complex (compared to simple collisional cloud models such as \citealt{dobbsw2021}) and as such, we might expect that locally filaments occur in environments with different densities and field strengths, leading to both parallel and perpendicular orientations, diluting the tendency for a clear preferred orientation.

\subsubsection{Magnetic criticality and magnetic energy}
\label{sec:magcrit}
We display the magnetic critical parameter $\mu_{\scriptscriptstyle B}$, which is the mass to flux ratio of a clump or cloud (subscript c refers to clump or cloud),
\begin{equation}
    \mu_{\scriptscriptstyle B} = \frac{M_{\scriptscriptstyle \textrm{c}}}{M_{\scriptscriptstyle \Phi}}, \ \ \ \ \ \text{where} \ \ \ \ \  M_{\scriptscriptstyle \Phi} \equiv \frac{0.12}{\sqrt{G}}\Phi_{\scriptscriptstyle B}, \ \ \ \ \ \text{and} \ \ \ \ \  \Phi_{B} = B\pi R^{2}_{\scriptscriptstyle \textrm{mag}},
    \label{eqn:critical}
\end{equation}
in the top panel of Figure \ref{fig:props_clumps} for clumps, and Figure~\ref{fig:props_clouds} for clouds. The magnetic field, $B$, is calculated from the average magnetic field strength of particles in a structure. $R_{\scriptscriptstyle \textrm{mag}}$ is the radius used to work out the flux, $\Phi_{\scriptscriptstyle B}$, through the centre of mass of the clump or cloud. This radius is calculated as \begin{equation}
    R_{\scriptscriptstyle \textrm{mag}} = \frac{R_{\scriptscriptstyle \textrm{COM}}}{3} \left(1+\frac{b}{a}+\frac{c}{a}\right). 
\end{equation}
where $R_{\scriptscriptstyle \textrm{COM}}$ is the radius of the clump calculated as the furthest particle away from the centre of mass (COM). $a$, $b$ and $c$ are the eigenvalues of the principle axes calculation mentioned early in this section. This equation takes into account the morphology of the clump, and avoids overestimating the flux surface for clumps that are not spherical, i.e. filaments and sheets. 
The mass to flux ratio, $\mu_{\scriptscriptstyle B}$, defines the magnetic criticality of a structure. For subcritical clouds, $\mu_{\scriptscriptstyle B}>1$, and the structure is magnetically supported whilst for supercritical clouds, $\mu_{\scriptscriptstyle B}<1$, and magnetic fields cannot support the structure against collapse. 

For the cloud models (Figure~\ref{fig:props_clumps}), Cloud5 is initially globally supercritical and Cloud50 globally subcritical. For clumps within the cloud though there is a spread in criticality, which increases with time. At early times, most clumps are subcritical in both models, but at later times there are also many supercritical clumps, and Cloud50 unexpectedly contains some strongly supercritical clouds. The strongly supercritical clumps we see in Cloud50 at 5 Myr are all small, dense, low mass clumps, which result from fragmentation and compression as feedback acts on the gas. In Cloud5 and Cloud0, the feedback simply disperses the gas and fewer such clumps survive. We also see that there is a much larger spread in $\mu_{\scriptscriptstyle B}$ for low mass clumps. Higher mass ($10^3-10^4$ M$_{\odot}$) clumps tend to have $\mu_{\scriptscriptstyle B}$ close to 1. 

For the Arm models (Figure \ref{fig:props_clouds}), most of the clouds are supercritical. The 
higher magnetic field strength models have more subcritical clouds as would be expected. 
There is some increase in spread in $\mu_{\scriptscriptstyle B}$ with time, but less than the Cloud models, and again the more massive clouds have the lowest degree of spread. The 
most supercritical clouds are similarly small, low mass structures. The development of small dense supercritical clumps is less evident in the Arm models, presumably because they are not resolved on these scales or the field strengths are too weak. 

In the lower panels of Figure \ref{fig:props_clumps} and \ref{fig:props_clouds} we show the ratio of the Alfvén speed to the sound speed of the gas, i.e. $\nu_{\scriptscriptstyle A}/\nu_{\scriptscriptstyle s}$, where $\nu_{\scriptscriptstyle A} = B_{\scriptscriptstyle\textrm{c}}/(4\pi\rho_{\scriptscriptstyle\textrm{c}})^{\scriptscriptstyle0.5}$ and $\nu_{\scriptscriptstyle s} = ((\gamma^{\scriptscriptstyle 2}-\gamma)u_{\scriptscriptstyle\textrm{c}})^{\scriptscriptstyle0.5}$, where $\gamma = 2.381$, $u_{\textrm{c}}$ is the internal energy, subscript c is cloud or clump. This ratio provides an indication of the dominant information speed within a structure. $\nu_{\scriptscriptstyle A}/\nu_{\scriptscriptstyle s}>1$ indicates magnetic information propagates faster than the gas information, suggesting a structure which is magnetically dominated.  
In Cloud5 and Cloud50 the structures are mostly supercritical and so correspondingly the dominating information speed is the sound speed. 
Again, the compression by feedback increases the density and temperature of the clumps, which decreases $\nu_{\scriptscriptstyle A}$ and increases $\nu_{\scriptscriptstyle s}$.
These kinematically driven clumps become increasingly supercritical since the magnetic field is much less responsive. 
The Arm models similarly show that most clumps are dominated by the sound speed. The ratio $\nu_{\scriptscriptstyle A}/\nu_{\scriptscriptstyle s}$ shows a similar trend to $\mu_{\scriptscriptstyle B}$, i.e. with increasing field strength, $\mu_{\scriptscriptstyle B}$ decreases and correspondingly $\nu_{\scriptscriptstyle A}/\nu_{\scriptscriptstyle s}$ increases.

We also computed the ratios of magnetic energy to the kinetic and gravitational energies. The ratios of the energies largely agree with the critical parameter, and most clouds have higher kinetic and gravitational energies compared to the magnetic energy. Other numerical work also finds that clouds tend to have greater gravitational energies (\citealt{hu2023,ganguly2023}), though Hu find the critical parameters can be subcritical when calculated using synthetic observations. \citet{ganguly2024} 
also find that the magnetic surface term can be a significant confining pressure, particularly for lower density clouds.

In Appendix \ref{append:sim_convergence} we test whether resolution changes the magnetic properties of our models, and compare the results from model Arm5 with a lower resolution version of Arm5. We do not see a significant difference between the magnetic properties of the clouds at lower resolution, indicating that our conclusions about magnetic fields are not strongly resolution dependent.

\subsection{Magnetic field density relation}
\label{sec:brho}
\begin{figure*}
    \centering
\includegraphics[width=140mm]{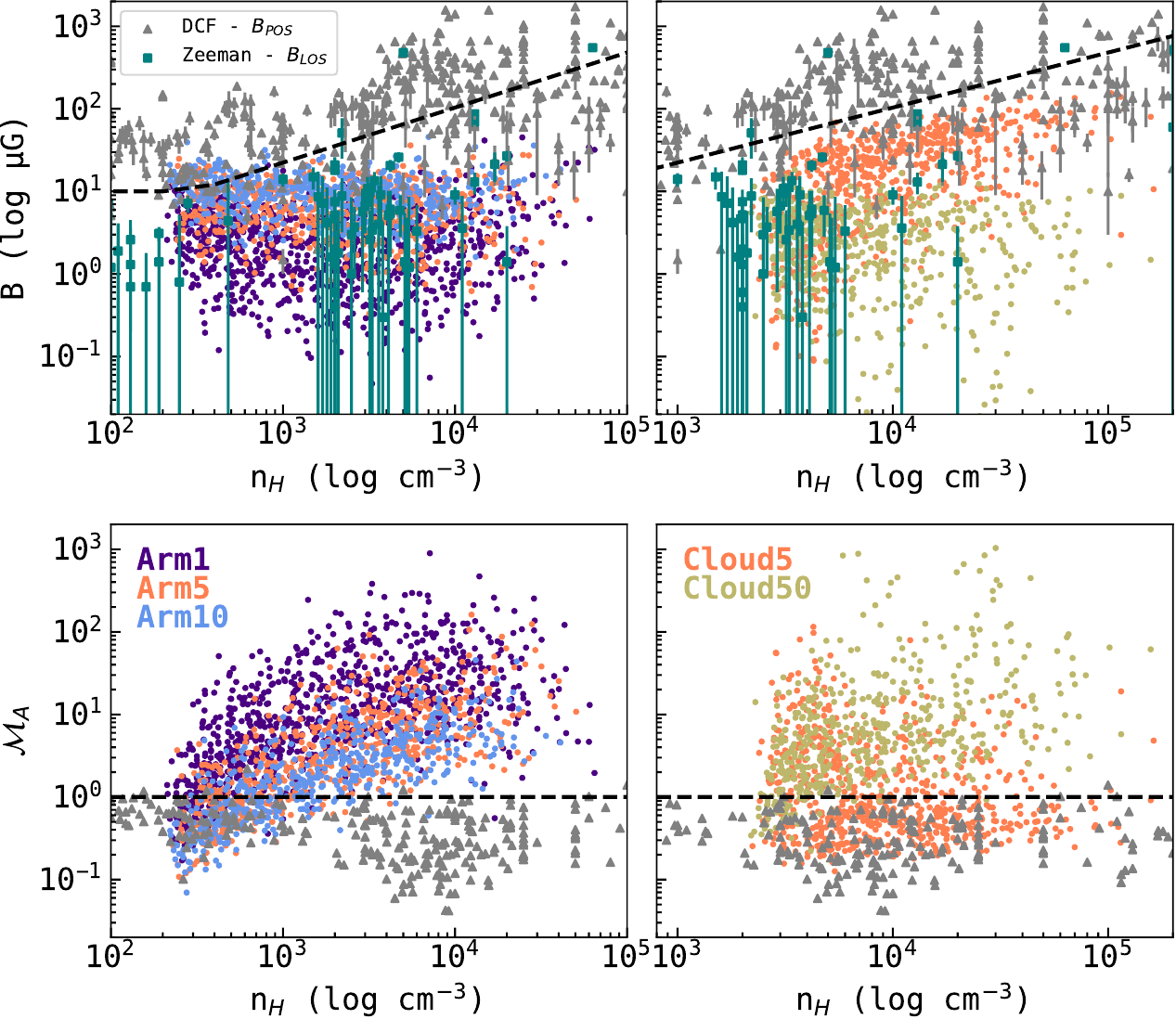}
    \caption{Magnetic field - density relation for Arm1, Arm5 and Arm10 (left panel) as well as Cloud5 and Cloud50 (right panel) in the top row. Alfvén Mach number against density for those same models in the bottom row. These are calculated at 6 Myrs for the Arm models and 5 Myrs for the cloud models. DCF and Zeeman measurements of $B_{\scriptscriptstyle\textrm{POS}}$ and $B_{\scriptscriptstyle\textrm{LOS}}$ are shown in the top panel \citep{crutcher2012,pattle2023}. In the bottom panel only DCF measurements are shown, which correspond to $\mathcal{M}_{\scriptscriptstyle A}=\sigma_{\scriptscriptstyle\theta}/Q$ (see \citealt{pattle2021apj} for details). The observations correspond to zoomed in regions from Figure~2a and 2e of \citet{pattle2023}.}
    \label{fig:obs}
\end{figure*}

The top Panel of Figure \ref{fig:obs} shows the magnetic field density relation $B-\rho$ for our Cloud and Arm models (in the right and left panels respectively). We compare this to the Crutcher relation which corresponds to a Bayesian model of the maximum magnetic field strength versus density obtained from
Zeeman measurements \citep{crutcher2010}.
\begin{equation}
    B = 
    \begin{cases}
    B_0  & n < n_0 \\
    B_0\left(\frac{n}{n_0}\right)^{k} & n > n_0,
    \end{cases}
\end{equation}
where $n_{\scriptscriptstyle 0} \sim 300$~cm$^{-3}$, $B_{\scriptscriptstyle 0} \sim 10$~$\mu$G and $\kappa = 2/3$ are estimated from Zeeman measurements of cloud structures 
\citep{crutcher2012}.
  We additionally plot the Davis-Chandrasekhar-Fermi or DCF observations (method - \citealt{Davis1951,Chandra1953}; data - \citealt{pattle2023} and references therein) and Zeeman measurements \citep{crutcher2010}. The DCF method finds $B_{\scriptscriptstyle\textrm{POS}}$, i.e. the position on the sky magnetic field and the Zeeman measurements calculate $B_{\scriptscriptstyle\textrm{LOS}}$, which is the line of sight magnetic field.
The DCF method estimates $B_{\scriptscriptstyle\textrm{POS}}$ through the dispersion in polarisation angle from dust emission or extinction. This can be used to estimate the disruption of the magnetic field by non-thermal gas motions, which can be a measure of the Alfvén Mac number $\mathcal{M}_{\scriptscriptstyle A}$. 
\begin{figure}
    \includegraphics[width=68mm]{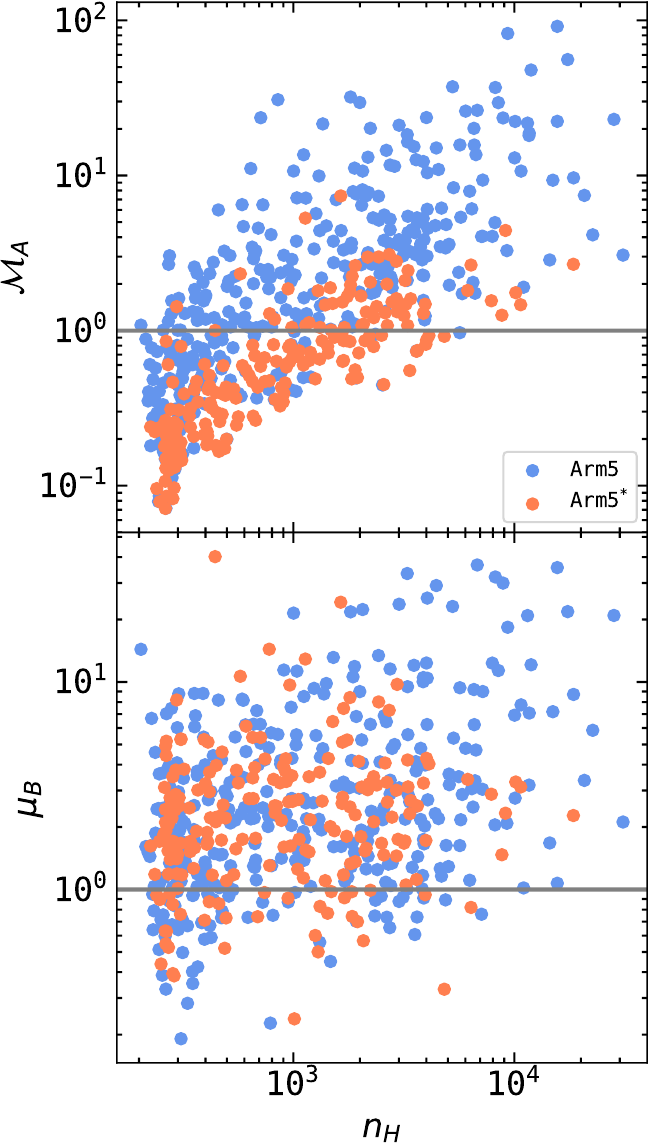}
    \caption{The Alfvén Mach number, $\mathcal{M}_{\scriptscriptstyle A}$, is shown in the top panel and the mass to flux ratio, $\mu_{\scriptscriptstyle B}$, in the bottom panel for clouds from models Arm5 (blue) and Arm5$^*$ (orange). Arm5$^*$ is the same as Arm5, except this model does not include stellar feedback. All properties are plotted with respect to number density $n_{\scriptscriptstyle H}$.}
    \label{fig:arm5_nofb_clouds}
\end{figure}
\begin{table*}
\begin{center}
\caption{Table of cluster masses for the Cloud models. The clusters are identified using HDBSCAN.}
\begin{tabular}{| c | c | c | c | c | c | c |}
 \hline
 Model & Cluster 1 (M$_{\odot}$) & Cluster 2 (M$_{\odot}$) & Cluster 3 (M$_{\odot}$) & Cluster 4 (M$_{\odot}$) & Cluster 5 (M$_{\odot}$) & Cluster 6 (M$_{\odot}$) \\ 
 \hline
 Cloud0 & 6704 & 2621 & 866 & 426 & 348 & 145 \\
 \hline
Cloud5 & 6735 & 2494 & 848 & 628 & 574 & 518 \\
 \hline
Cloud50  & 3651 & 1420 & 912 & 852 & - & - \\
 \hline
\end{tabular}  
\label{tab:clusters}
\end{center}
\end{table*}
The Arm and Cloud models roughly span the density range $n_{\scriptscriptstyle H}\sim 10^{2} - 10^{5}$ cm$^{-3}$ and show better agreement with the Zeeman measurements compared to the DCF measurements at these densities. The Arm models 
find $B$ is fairly constant with density. For the Arm10 model, $B$ remains effectively constant across all densities, though for the Arm1 and Arm5 models, $B$ starts to increase slightly with density at around $n_{\scriptscriptstyle H}\gtrsim3\times10^3$~cm$^{-3}$. 
The Cloud5 model shows an increase in $B$ with density, which by eye matches the data well. We fit an exponent of $\kappa = 0.861$ to the simulated data. For the Cloud50 model however, the magnetic field is fairly constant and does not match the observations that well. Overall the higher magnetic field strength in the Arm10 and Cloud50 models seems to impede the formation of dense, strongly magnetic clouds and clumps.
\subsection{Turbulence and magnetic fields}
Combining the Alfvén speed, $\nu_{\scriptscriptstyle A}$ of a structure along with the non thermal velocity dispersion, $\sigma_{\scriptscriptstyle v}$ of the gas gives us the Alfvén Mach number, $\mathcal{M}_A = \sigma_v/v_A$. This indicates the level of magnetic turbulence within a structure, $\mathcal{M}_{\scriptscriptstyle A}<1$ is sub-Alfvénic and $\mathcal{M}_{\scriptscriptstyle A}>1$ is super-Alfvénic turbulence. In our simulations we compute $\sigma_v$ and $\nu_{\scriptscriptstyle A}$ directly from the gas particles within a given structure.  

In the bottom panel of Figure \ref{fig:obs} we show the Alfvén Mach numbers for the Arm and Cloud models compared to DCF measurements (described in §\ref{sec:brho}). 
\citet{pattle2023} explains that the DCF measurements of $\mathcal{M}_{\scriptscriptstyle A}$ assume sub-Alfvénic turbulence within the gas, 
which is why these observations generally find $\mathcal{M}_{\scriptscriptstyle A} < 1$. Zeeman measurements span lower magnetic fields which would correspond to higher Alfvén Mach numbers more in line with our models. 
Sub-Alfvénic turbulence is common in our Arm and Cloud models but we also observe significant super-Alfvénic turbulence within clouds and clumps. The mean Mach numbers within the Arm models Arm1, Arm5 and Arm10 are $\mathcal{M}_{\scriptscriptstyle A} = $ 14.0, 4.93 and 2.05 respectively, i.e. super-Alfvénic. The median values for the Arm models are super-Alfvénic for Arm1 and roughly trans-Alfvénic for Arm5 and Arm10. Our cloud models reflect this behaviour with mean values super-Alfvénic and median values trans-Alfvénic, but Cloud5 shows more sub-Alfvénic clumps compared to Cloud50.

We investigate the relationship between feedback and magnetic fields by calculating the Alfvén Mach number and criticality of cloud structures within Arm5 and a new model Arm5$^{*}$. This new model has the same initial conditions as Arm5 but evolves without stellar feedback. In the top panel of Figure \ref{fig:arm5_nofb_clouds} we show the Alfvén Mach Number $\mathcal{M}_{\scriptscriptstyle A}$. We find significantly more super-Alfvénic clouds, as would be expected for feedback driving turbulence, in Arm5 compared to Arm5$^{*}$, as well as more higher density clouds. The critical parameter for clouds in both models is shown in the bottom panel. For structures of the same density we find the criticality of clouds to be roughly the same in Arm5 and Arm5$^{*}$, indicating that feedback is not obviously effecting magnetic criticality.
\begin{figure}
    \centering
    \includegraphics[width=70mm]{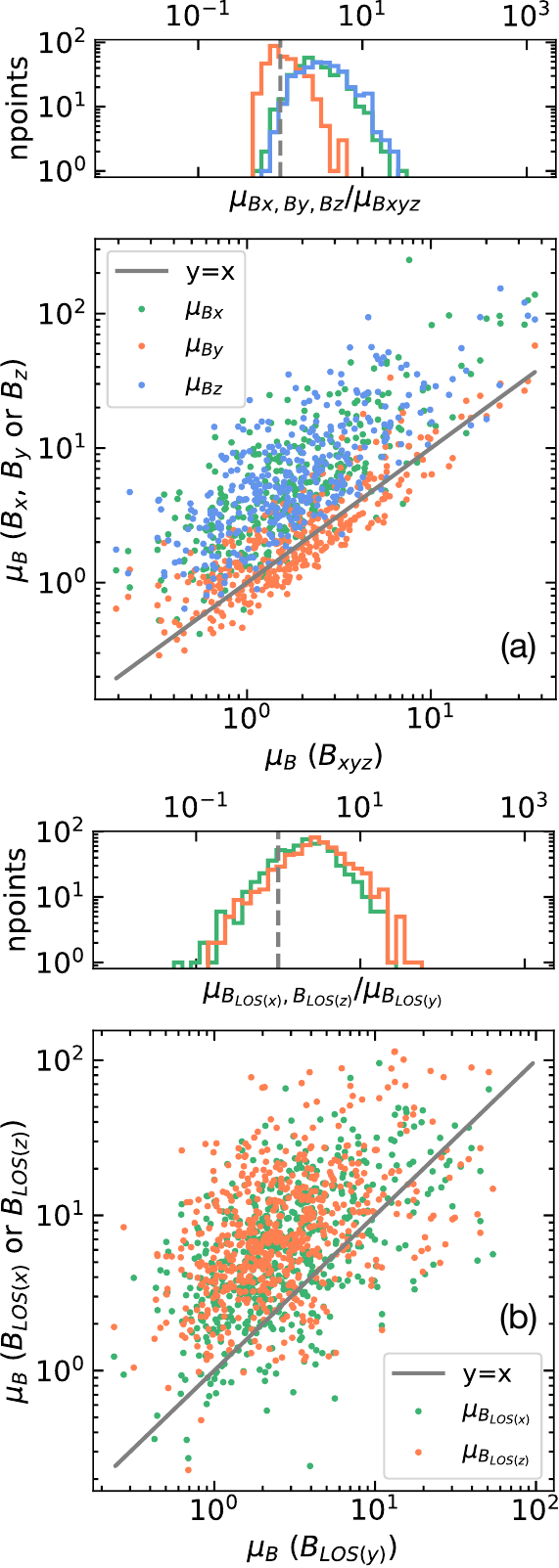}
    \caption{(a) - The magnetic critical parameter $\mu_{\scriptscriptstyle B}$ of clouds in Arm5 is plotted, computed using using $B_{\scriptscriptstyle x}$, $B_{\scriptscriptstyle y}$ or $B_{\scriptscriptstyle z}$ only compared to 3D $B_{\scriptscriptstyle xyz}$. (b) - $\mu_{\scriptscriptstyle B}$ calculated using $B_{\scriptscriptstyle\textrm{LOS(x)}}$ or $B_{\scriptscriptstyle\textrm{LOS(z)}}$ compared to $B_{\scriptscriptstyle\textrm{LOS(y)}}$ (see text). These measures of $\mu_{\scriptscriptstyle B}$ are calculated using the column density along the line rather than the flux, but give similar values to the flux method.}
    \label{fig:mu_bxyz}
\end{figure}   
\subsection{Clusters and magnetic fields}
For the Cloud models, the resolution is sufficiently high that we are able to identify clusters from the sink particles formed in the simulation. We use HDBSCAN \citep{campello2013} to identify clusters, since this is used commonly by observers \citep{joncour2020}. We also used this in previous work \citep{ali2023}. We set the minimum cluster size \textit{min\_cluster\_size = 15}, which sets the minimum number of particles in a cluster. We define \textit{min\_samples = 3}, which is a measure of how conservative the clustering algorithm is, and how many points remain as noise.

From Figure \ref{fig:cd_all}, we see by eye that the clusters seem less extensive in Cloud50, whereas for Cloud0 and Cloud5 the distribution of sink particles looks similar. We show the number of clusters, and the cluster masses, from the clusters identified with HDBSCAN in Table \ref{tab:clusters}. The number of clusters is the same for Cloud0 and Cloud5, and the masses of the clusters fairly similar (though the masses of the smallest clusters tend to be a little higher in Cloud5).  By contrast, for Cloud50, there are only 4 clusters, instead of 6. The masses of the two most massive clusters are reduced by about 40\%.
\section{Discussion}
\label{sec:disc}
\subsection{Directional dependence of $\mu_{\scriptscriptstyle B}$}
Our simulations provide us with the full 3D magnetic field properties. 
DCF methods 
only recover 2D magnetic field information, i.e. the position on the sky magnetic field $B_{\scriptscriptstyle\textrm{POS}}$. The most direct method,
Zeeman splitting,
measures the magnetic field strength $B_{\scriptscriptstyle\textrm{LOS}}$ in the direction of the line of sight only. 
We consider how the magnetic critical parameter depends on $B$, assuming that we only have the magnetic field information in a particular direction, and using a line of sight method. 

We first see how magnetic field direction affects $\mu_{\scriptscriptstyle B}$, as calculated in equation \ref{eqn:critical}. For simplicity, we take $B_x$, $B_y$, $B_z$, for equation \ref{eqn:critical}, noting that for our region, the magnetic field is most aligned with the $y$ direction.
In panel (a) of Figure \ref{fig:mu_bxyz} we show the critical mass to flux ratio for cloud structures in Arm5, computed with the full 3D field strength $B_{\scriptscriptstyle\textrm{xyz}}$ and single field directions $B_{\scriptscriptstyle\textrm{x}}$, $B_{\scriptscriptstyle\textrm{y}}$ or $B_{\scriptscriptstyle\textrm{z}}$.  We observe that $\mu_{\scriptscriptstyle B_{\scriptscriptstyle\textrm{y}}}$ produces a similar mass to flux ratio to $\mu_{\scriptscriptstyle B_{\scriptscriptstyle\textrm{xyz}}}$ whereas $\mu_{\scriptscriptstyle B_{\scriptscriptstyle\textrm{x}}}$ and $\mu_{\scriptscriptstyle B_{\scriptscriptstyle\textrm{z}}}$ 
produce significantly larger critical parameters, i.e. more supercritical. We find $<\mu_{B_{\scriptscriptstyle xyz}}/\mu_{B_{\scriptscriptstyle x}}> = 3.7$ and $<\mu_{B_{\scriptscriptstyle xyz}}/\mu_{B_{\scriptscriptstyle z}}> = 4.3$ corresponding to the peaks of the histogram in the top panel of Figure \ref{fig:mu_bxyz}.

The higher $\mu_{\scriptscriptstyle B_{\textrm{x}}}$, $\mu_{\scriptscriptstyle B_{\textrm{z}}}$ 
are due to lower magnetic field strengths in these directions, compared to the $y$ direction. This is because the global field is most aligned with the $y$ direction. We can estimate the predicted directional dependence as follows. Observations find that the random component of the field is about 1/3 of the ordered component \citep{fletcher2010}. If we are looking in the direction of the ordered component, which will tend to be the rotation of the galaxy or along the spiral arms, we will see a maximum difference of a factor of 3 between the magnetic field strength, or critical parameter, compared to perpendicular directions. If we look in the plane of the disc but perpendicular to the ordered field, we would expect minimal directional dependence. But overall we would expect direction to contribute a factor of 3 uncertainty.

In panel (b) we computed the critical parameters using a 
column density method, again for the Arm5 model. We take lines of sight over a grid (with spacing 1 pc), and for any line of sight which passes within a smoothing length of particles in the cloud, we compute the column density along the line of sight from material in the cloud, and the mean magnetic field over those constituent particles. The magnetic criticality is then calculated as
\begin{equation}
\mu_{B_{\scriptscriptstyle\textrm{LOS}}}=\frac{\mu_{\scriptscriptstyle H} N}{B},    
\end{equation}
where $\mu_{\scriptscriptstyle H}$ is the mean molecular weight and $N$ is the column density \citep{crutcher1999}.
This is more analogous to Zeeman observations. We compare the LOS derived critical parameters in the y direction to the other weaker field strength components.
This gives
$<\mu_{B_{\scriptscriptstyle\textrm{LOS(y)}}}/\mu_{B_{\scriptscriptstyle\textrm{LOS(x)}}}> = 3.1$ and $<\mu_{B_{\scriptscriptstyle\textrm{LOS(y)}}}/\mu_{B_{\scriptscriptstyle\textrm{LOS(z)}}}> = 4.1$. So both the values for $\mu$, and the relative differences between directions, are similar for both types of calculations of $\mu$.

\subsection{Magnetic fields}
We have run models with varying magnetic field strengths. Morphologically, the subcritical Cloud50 model shows a large difference in the structure of the gas, and the reduced effect of feedback, compared with the purely hydrodynamical model. The magnetic field also has a much greater impact on star formation, reducing the amount of star formation by as much as 50\%, and the mass of clusters by a factor of around a half. For Cloud5, the morphology of the gas is more similar to the hydrodynamical model, the clusters are similar and the star formation rate reduced by a smaller amount. So in this case, the behaviour of Cloud5 seems more representative of the weak field scenario, whilst Cloud50 is more representative of the strong field scenario. 

The main diagnostic for comparing with observations we have shown is the B-$\rho$ relation. We see that for the strongest field case, Cloud50, the results do not match the observations particularly well, in that $B$ is flat versus density instead of increasing, whereas Cloud5 matches much better. We further tested this by running a model Arm50 (with a 50~$\mu$G initial field), and this model also produced points that disagreed with the observations, showing higher $B$ values which are flat with density and cross the Crutcher relation. For both of these models with 50~$\mu$G fields, the magnetic fields are strongly preventing gravitational collapse. As discussed in \citet{crutcher2010}, the transition from a flat relation to a $B\propto\rho^{2/3}$ relation is expected to occur when clouds can contract via gravity. Thus it is not surprising that these stronger field models show no upturn in $B$.
The Arm50 model also looked morphologically dissimilar to what we would expect, with filaments forming perpendicular, rather than parallel to the arm (extending the trend we see in Figure \ref{fig:angles_clumps}). For larger, lower density, arm scales we would expect to see filaments more parallel to the magnetic field, and both roughly aligned with the spiral arms, like Arm1 and a lesser extent Arm5. So overall, our models with weaker fields are more consistent with observations. A caveat to our models is that we do not include ambipolar diffusion, which could allow gravitational collapse even in the strong field cases, although for our Arm models at least, we would not expect ambipolar diffusion to be relevant on these scales.

Observationally, we see from the $B-\rho$ relation that the DCF measurements tend to be higher than the Zeeman measurements. The DCF method is more uncertain compared to the Zeeman method, since the latter measures the field directly rather than inferring the field from the dispersion of polarisation vectors. Our models show magnetic field strengths which agree better with the Zeeman, rather than the DCF measurements, and thus we find that the Zeeman measurements are a better match to the models. As mentioned above, for higher magnetic fields, our models still don’t fit the DCF measurements, because the B-$\rho$ relation is too flat.

\section{Conclusions}
\label{sec:conclusions}
We have presented 3D smooth particle magnetohydrodynamic simulations of star formation on kpc and pc scales, following the evolution of a section of spiral arm and a single cloud structure. Listed below is a summary of our main conclusions from the results of our simulations.
\begin{itemize}
    \item[(1)] We 
    find that magnetic fields cause smoothing of dense structures leading to smoother feedback shells and filamentary networks compared to our purely hydrodynamic models.
    \item[(2)] Magnetic fields suppress star formation rates on both kpc and pc scales. The strongest suppression of the SFR on kpc scales occurs in the highest strength model (Arm10), where the SFR is reduced 
    by roughly 30\%, comparable to previous results (e.g. \citealt{wibking2023}). On pc scales our strong field model (with $B_{\scriptscriptstyle 0}=50~$ $\mu$G) shows suppression of up to $\sim$50\%. The decrease in star formation leads to a reduction in the number of clusters from 6 to 4, and decrease in mass of the most massive cluster by 40\%, in the strongest field model.
    However, with a more moderate field ($B_{\scriptscriptstyle 0}=5~$ $\mu$G), the number and masses of clusters are similar to the hydrodynamic case.  
    \item[(3)] With less star formation and therefore feedback, there is also less gas dispersal in our Cloud model with the strongest field. 
    This likely prolongs the cloud's lifetime and duration of star formation compared to our hydrodynamical model. 
    \item[(4)] Cloud structures identified in the Arm models tend to be more filament like and aligned parallel to their magnetic fields in low magnetic field strength models and perpendicularly in higher magnetic field strength models. The transition in orientations at higher 
    magnetic field strengths is consistent with previous work \citep{solerh2017,seifried2020,dobbsw2021,barretomota2021,mazzei2023}. We do not find a transition between the flow and magnetic field direction, both Arm1 and Arm10 show flow aligned to the direction of the magnetic field. Filamentary structures within our cloud models do not show a clear alignment preference, contrary to similar scale structures observed inside molecular clouds (e.g. \citealt{planckcolab2016,solera2017,alina2019,lee2021}). 
    \item[(5)] 
    We see 
    a large spread in criticality for both clumps and clouds throughout the simulations.
    We surprisingly see more supercritical low mass clumps at later times, particularly in Cloud50. We attribute these to feedback sweeping gas into dense shells, with small clumps occurring as the shell fragments.  
    \item[(6)] We find the critical parameter has a directional dependence. We compared calculations of the critical parameter using only one component of the magnetic field, $B_{\scriptscriptstyle x}$, $B_{\scriptscriptstyle y}$ and $B_{\scriptscriptstyle z}$, giving us $\mu_{\scriptscriptstyle B_{\scriptscriptstyle x}}$, $\mu_{\scriptscriptstyle B_{\scriptscriptstyle y}}$ and $\mu_{\scriptscriptstyle B_{\scriptscriptstyle z}}$. This was compared to $\mu_{\scriptscriptstyle B_{\scriptscriptstyle xyz}}$ where $B_{\scriptscriptstyle xyz}$ is the total 3D magnetic field strength. We find $\mu_{\scriptscriptstyle B_{\scriptscriptstyle x}}$ and $\mu_{\scriptscriptstyle B_{\scriptscriptstyle z}}$ overestimate the critical parameter by roughly 3-4 times compared to $\mu_{\scriptscriptstyle B_{\scriptscriptstyle xyz}}$,
    whereas the component $\mu_{\scriptscriptstyle B_{\scriptscriptstyle y}}$ produces effectively the same values as $\mu_{\scriptscriptstyle B_{\scriptscriptstyle xyz}}$.
    This level of variation might be expected if the random component of the galactic magnetic field is of order 1/3 the ordered component. We also compared calculating the critical parameter using the line of sight column density versus using the flux and found the methods produced similar results.

    \item[(7)] We determine the $B-\rho$ relation for our models, finding that the magnetic field measurements for our clumps and clouds agree better with
    Zeeman observations compared to DCF measurements. Our stronger field models ($B_{\scriptscriptstyle 0}=50~$ $\mu$G and 10~$\mu$G) show a flat $B-\rho$ relation, whereas Cloud5 ($B_{\scriptscriptstyle 0}=5~$~$\mu$G) and to a lesser extend Arm1 and Arm5 show an increasing $B-\rho$ relationship at higher densities. The change from flat to increasing $B-\rho$ is expected to coincide with gravitational contraction \citep{crutcher2010}, which may explain why our stronger field models, where star formation is significantly impeded, exhibit a flat $B-\rho$ relationship.
    \item[(8)] DCF measurements predict larger magnetic fields and so greater Alfvén speeds, recovering on average sub-Alfvénic ($\mathcal{M}_{\scriptscriptstyle A}$) turbulence on density scales $n_{\scriptscriptstyle H}\leq10^6$ cm$^{-3}$. Our models show significant numbers of super-Alfvénic structures on both arm and cloud scales for densities $n_{\scriptscriptstyle H}\sim10^2-10^5$.
    \item[(9)] Finally we report that  feedback does not effect the magnetic criticality of clouds. Instead we observe more super-Alfvénic clouds in magnetic models with feedback compared to magnetic models without feedback. 
\end{itemize}

As discussed previously, changing the resolution does not seem to affect the magnetic properties, but there are a few further caveats. One is that we start with a uniform field rather than taking a field from a global galaxy simulation. As we don’t have a suitable MHD galaxy simulation, we cannot test directly what difference this makes. However in Appendix \ref{append:initial_mtof} we consider the magnetic flux ratios over the course of the Arm simulations, and show they are not hugely different at later times, the main change is that the spread in the flux ratios increases. A further caveat is that when we increase the resolution we do not follow turbulence on the smallest scales, which may influence when and where star formation occurs. We tested this using a low resolution simulation equivalent to Arm5LR, but where we inject turbulence on small scales. We do this by adding a velocity dispersion when we split the particles, by sampling from a Gaussian distribution of width equal to the velocity dispersion expected from an empirical Larson relation. We find little difference in morphology and number of sinks formed compared to Arm5LR.

\section*{Acknowledgements} 
We would like to acknowledge James Wurster for his contribution to this work, in particular advice regarding using a boundary medium. We additionally acknowledge Kate Pattle for sharing the DCF data which enabled us to compare our results to observations. As well as this we would like to acknowledge the referee for their helpful comments and suggestions. Lastly we acknowledge Tim Naylor for our useful discussions.
All authors are funded by the European Research Council H2020-EU.1.1 titled the ICYBOB project, grant number 818940. Our simulations were performed with the DiRAC DIaL system, facilitated by the University of Leicester IT Services, forming part of the STFC DiRAC HPC Facility (www.dirac.ac.uk). The equipment was funded by BEIS capital funding via STFC capital grants ST/K000373/1 and ST/R002363/1 and STFC DiRAC Operations grant ST/K001014/1. DiRAC is part of the National E-Infrastructure.

\section*{Data availability statement}
Upon reasonable request to the author, all data in this work will be shared.



\bibliographystyle{mnras}
\bibliography{refs} 




\appendix
\section{Resolution test}
\label{append:sim_convergence}
\begin{figure}
    \centering
    \includegraphics[width=60mm]{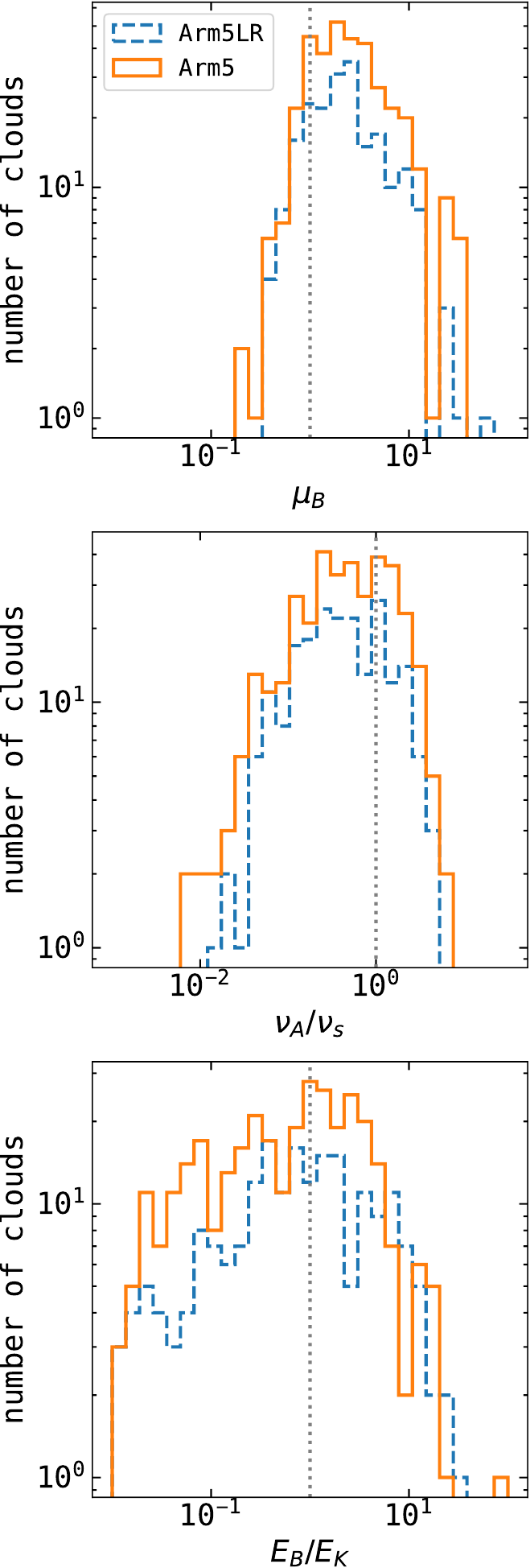}
    \caption{Histograms of mass to flux ratio $\mu_{B}$, the ratio of Alfv\'en speed to sound speed $\nu_{A}/\nu_{s}$ and the ratio of magnetic to kinetic energy of identified cloud structures in Arm5 (orange line) and new model Arm5LR (blue dashed line), calculated at 4 Myrs.}
    \label{fig:LRtest}
\end{figure}

We produced a new model named Arm5LR which has the same initial conditions as Arm5 but has a lower resolution of 10.5~M$_{\odot}$ per particle. Figure \ref{fig:LRtest} shows a comparison of magnetic properties of clouds identified in both Arm5 and Arm5LR. We find that all magnetic properties are similar between Arm5 and the lower resolution model Arm5LR. 

\section{Classifying structures}
\begin{figure*}
    \centering
    \includegraphics[width=140mm]{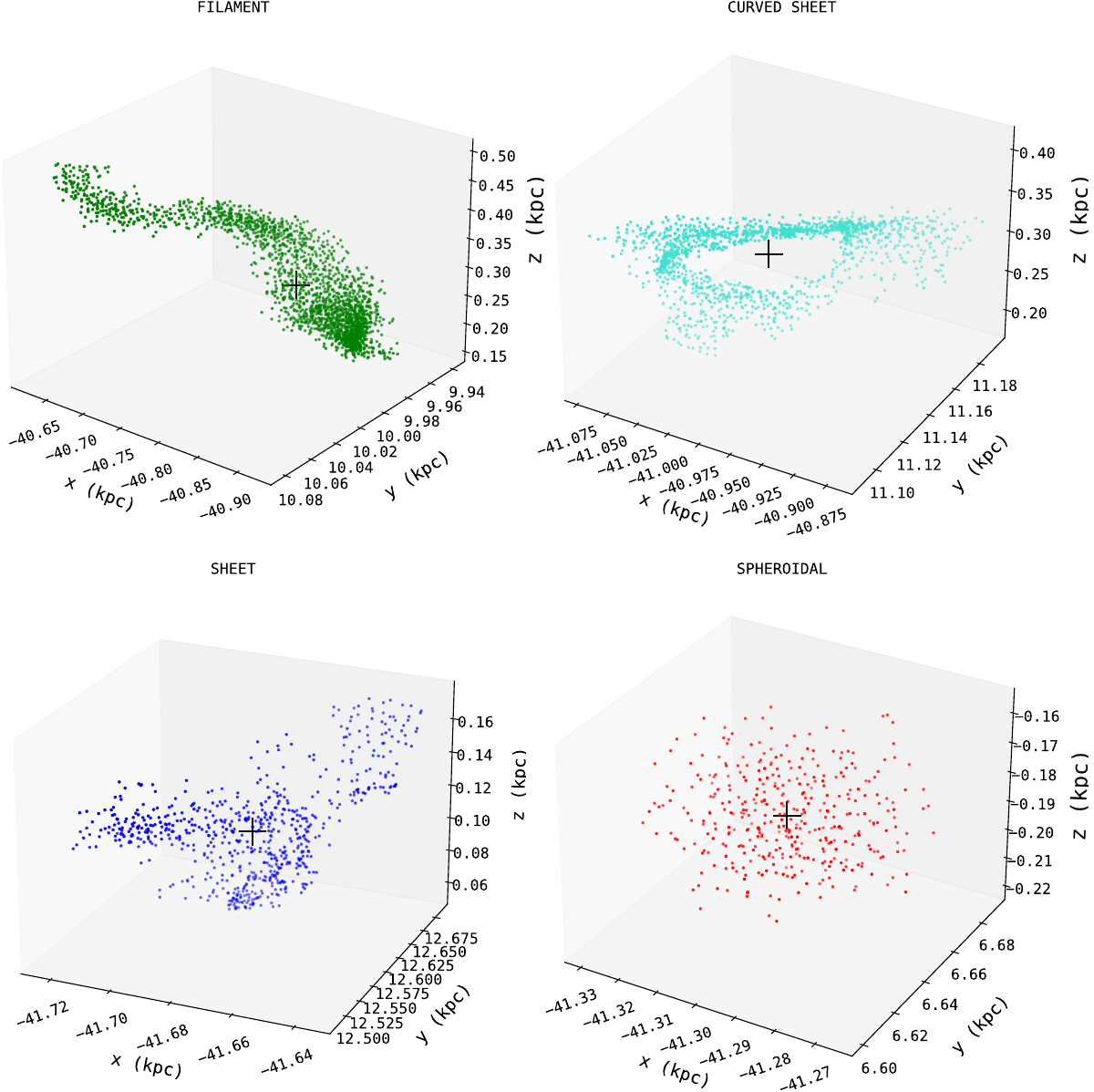}
    \caption{Structures classified as filaments, sheets, curved sheets and spheroidal. Circle points are particles that make up these structures and the cross is the centre of mass of the structure. These are examples of cloud structures from our Arm10 model.}
    \label{fig:appendix_classify}
\end{figure*}
We present examples of the classified structures mentioned in section \ref{sec:cl_props}. Figure \ref{fig:appendix_classify} shows cloud structures that correspond to filaments, sheets, curved sheets and spheroids. 
\begin{figure*}
    \centering
    \includegraphics[width=140mm]{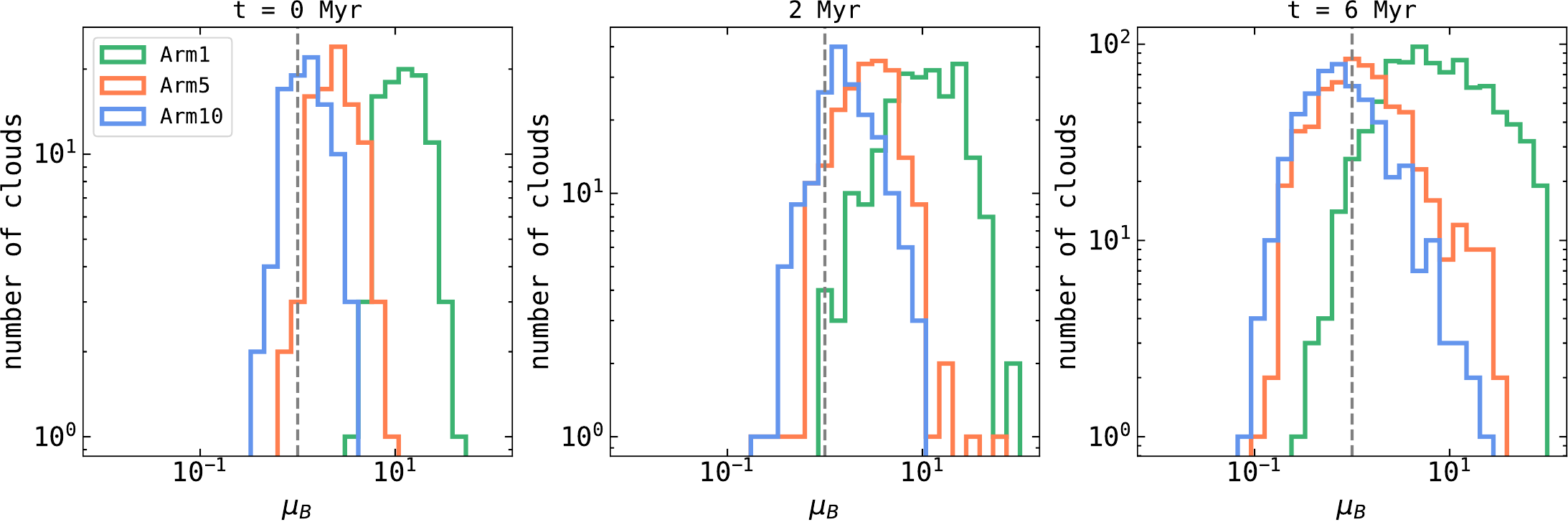}
    \caption{Mass to flux ratios $\mu_{B}$ of clouds in Arm1 (green), Arm5 (orange) and Arm10 (blue) at times t = 0, 2 and 5~Myrs.}
    \label{fig:mtof_tests}
\end{figure*}
\section{initial distribution of magnetic properties}
\label{append:initial_mtof}

The initial conditions of the Arm and Cloud models are extracted from non-MHD galaxy models. 
For our Arm and Cloud models the magnetic field in our initial conditions are generated by setting up a toroidal magnetic field. 
We compare the mass to flux ratio of clouds identified at t = 0, so from the initial conditions, to the rough middle and end of the simulation (4 and 6 Myrs respectively). These are shown in Figure \ref{fig:mtof_tests},  and we find the distributions of mass to flux ratios exhibit similar average values from t = 0 to t = 6 Myrs, although the spread in the values increases with time.


\bsp	
\label{lastpage}
\end{document}